\documentclass[12pt, letterpaper]{article}
\usepackage[top=1in, bottom=1.5in, left=1in, right=1in]{geometry}
\usepackage[utf8]{inputenc}
\usepackage{booktabs}
\usepackage{hyperref}
\usepackage{natbib}
\usepackage{graphicx}

\title{A Higher Cadence Subsurvey Located in the Galactic Plane}
\author{M.B.~Lund, K.G.~Stassun, J.~Farihi, E.~Agol, M.~Rabus, A.~Shporer, and K.J.~Bell \\with the support of the LSST Transient and Variable Stars Collaboration}
\date{Nov 2018}

\begin{document}

\maketitle

\begin{abstract}
Presently, the Galactic plane receives relatively few observations compared to most of the LSST footprint. While this may address static science, the plane will also represent the highest density of variable Galactic sources. The proper characterization of variability of these sources will benefit greatly from observations at a higher cadence.
\end{abstract}

\section{White Paper Information}
%Please provide contact information (name and email address) of the appropriate author(s) for this white paper.
%In addition, please provide the following categorization for your white paper:
\begin{enumerate} 
\item {\bf Science Category:} Transient and Variable Science %\emph{which of the four main LSST science themes are addressed? Are there other
%science programs addressed by this white paper?}
\item {\bf Survey Type Category:} Mini survey %{please choose one of the following possibilities: the main `wide-fast-deep'
   %survey, mini survey, Deep Drilling field, Target of Opportunity observation, Other (provide details).} 
\item {\bf Observing Strategy Category:} an integrated program with science that hinges on the combination of pointing and detailed 
	observing strategy
%please choose one of the following possibilities: 
    %\begin{itemize} 
    % \item a specific observing strategy to enable specific time domain science, 
	%that is relatively agnostic to where the telescope is pointed (e.g., a science case enabled 
	%by relatively deep precise time-resolved multi-color photometry). 
    % \item a specific pointing or set of pointings that is (relatively) agnostic of the detailed observing %
	%strategy or cadence, (e.g., a science case enabled by very deep precise multi-color 
	%photometry)
    %  \item an integrated program with science that hinges on the combination of pointing and detailed 
	%observing strategy (e.g., search for variable stars in the 
	%LMC/SMC). 
    %   \item other category (please describe).
    %\end{itemize}  
\end{enumerate}  

\noindent{\bf Authors:}\\
M.B.~Lund, Vanderbilt University, Michael.B.Lund@Vanderbilt.edu \\
K.G.~Stassun, Vanderbilt University, keivan.stassun@vanderbilt.edu \\
J.~Farihi, University College London, j.farihi@ucl.ac.uk\\
E.~Agol, University of Washington, agol@uw.edu\\
M.~Rabus, Pontificia Universidad Cat\'olica de Chile, markus.rabus@gmail.com \\
A.~Shporer, MIT, shporer@mit.edu\\
K.J.~Bell, Max Planck Institute for Solar System Research, bell@mps.mpg.de
\clearpage

\section{Scientific Motivation}
Studies of variability in stars generally require making trade-offs between survey area, cadence, and depth. For many surveys thus far, the largest sacrifices have typically been made in survey depth, and so the focus has been on brighter stars. This creates a bias towards stars that are either close and/or massive, limiting the study that can be carried out on certain stellar populations, such as intrinsically faint stars (e.g.\ white dwarfs, ultracool dwarfs) or intrinsically distant populations (e.g.\ solar-mass stars in bulge, globular clusters). The light-gathering capacity of LSST provides a unique opportunity in this regard, as this allows for a faint  survey, over a large area of the sky, at a suitable cadence.
\subsection{Transiting Planets}
Understanding planet formation has been one of the prominent goals in modern astrophysics, and a key component will be understanding how the stellar environment impacts planetary formation. This requires understanding planetary occurrence rates that go beyond the current base of knowledge -- primarily solar-mass stars in the approximate solar neighborhood.

Initial work has already demonstrated that transiting exoplanets can be detected with LSST \citep{Lund2015}. Further work focused on LSST's ability to find planets around a solar-mass star as a function of period and radius, showing that the present wide-fast-deep cadence was well-suited primarily for Hot Jupiters in very short orbits, and that the higher deep-drilling cadence could be used to detect a significant portion of exoplanets out to a period of 20 days \citep{Jacklin2015}. The stellar populations that LSST would be exploring have not been as thoroughly studied by other transit surveys, such as stars in the Galactic Bulge and Plane or red dwarfs. Understanding planet occurrence rates for these different populations will help explore the impact of stellar environment on planetary occurrence rate.
%An extension of this paper followed and demonstrated that LSST would only need 1-2 years of the deep-drilling cadence to detect a significant fraction of transiting hot Jupiters around G-dwarfs and hot Neptunes around K-dwarfs \citep{Jacklin2017}.

\subsection{Transits of White Dwarfs}
It is clear through multiple lines of evidence that planetary systems survive, at least in part, the post-main sequence evolutionary stages of their host star (see reviews \citealt{jura2014,farihi2016}).  The most sensitive of these metrics is atmospheric metal pollution that is seen in at least 30\% of all white dwarfs \citep{koester2014}, indicating these evolved planetary systems are common.  The potential for transiting planets around white dwarfs represents a unique and novel subset of transiting planets, but the occurrence rate of planetary-mass objects orbiting white dwarfs remains  poorly constrained \citep{vanSluijs2018}.  However, the 
\emph{K2} and ground-based detection of debris clouds transiting the polluted white dwarf WD 1145+017  \citep{Vanderburg2015} represented a significant first and suggests individual objects in orbit.   

The chance of an object transiting a white dwarf is $\sim100\times$ smaller than for planets around main-sequence stars of comparable masses.  And while white dwarfs are more common that G-type stars, they are typically $10^3-10^4$ times fainter than the Sun and become numerous at $m > 15$\,mag; thus beyond the range of previous surveys. 

\begin{figure}[!ht]
    \centering
    \includegraphics[width=11cm]{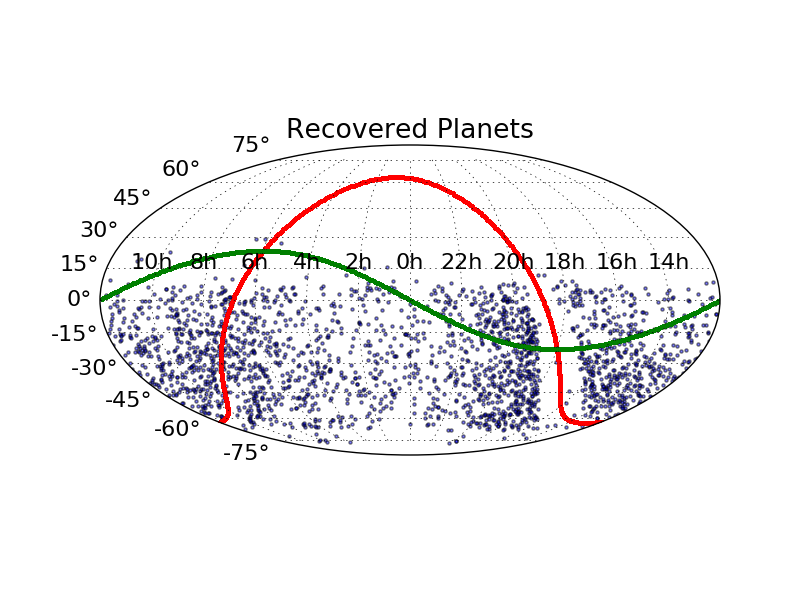}
    \caption{Map of recovered white dwarfs from simulated light curves published in \citet{Lund2018}. The Galactic plane accounts for 75\% of the white dwarfs in the simulation but are substantially less likely to be recovered. The impact of number of visits is explored in Figure~\ref{fig:WD_observations}.}
    \label{fig:WD_map}
\end{figure}

However, LSST's fainter magnitude range means that instead of $\sim1000$ white dwarfs monitored, LSST can observe millions and thus increase chances of observing objects in transit by orders of magnitude, an opportunity discussed in \citet{Agol2011b,Agol2011a}. Recent work has expanded on this, where \citet{Cortes2018} use the significance of transit signals in simulated observations to highlight that tens to thousands of Ceres-sized objects could be discovered with LSST, as well as hundreds of earth-sized planets in the habitable zone of those white dwarfs. A different approach taken in \citet{Lund2018} was to simulate detection of Earth-sized planets with a specialized algorithm, but similarly finds that hundreds of exoplanets can be directly discovered from simulated light curves. While most white dwarfs simulated in \citet{Lund2018} are located in the Galactic plane, very few of these are detected (see Figure~\ref{fig:WD_map}), a result of the low number of observations in the plane.

One complication for planets is the false-alarm rate, and follow-up resources required to confirm transiting planet candidates.  However, other genuine variability signals are of astrophysical interest in their own right, such as white dwarf binaries that might comprise Type Ia supernova progenitors, or white dwarfs occulted by debris disks. The particular topic of pulsations of white dwarfs with LSST has also already been highlighted \citep{Bell2015}.

This science case is also discussed in a proposal for LSST Wide-Fast-Deep observations in the white paper Silvotti et al.

\subsection{Other Science Cases}
The study of most categories of Galactic variables would benefit from increased observation in the plane as the number of observed objects would likely greatly increase, and are presented in other white papers (Street et al and Strader et al). These cases include numerous categories of stellar variability and microlensing events for a range of lensing masses, from planetary-mass objects to intermediate-mass black holes. Static science in the Galactic plane is discussed in the white paper Gonzalez et al.
%\begin{footnotesize}
%{\it Describe the scientific justification for this white paper in the context of your field, as well as the importance to the general program of astronomy, including the relevance over the next decade. Describe other relevant data, and justify why LSST is the best facility for these observations. (Limit: 2 pages + 1 page for figures.)}
%\end{footnotesize}
\clearpage

\begin{figure}[!ht]
    \centering
    \includegraphics[width=9cm]{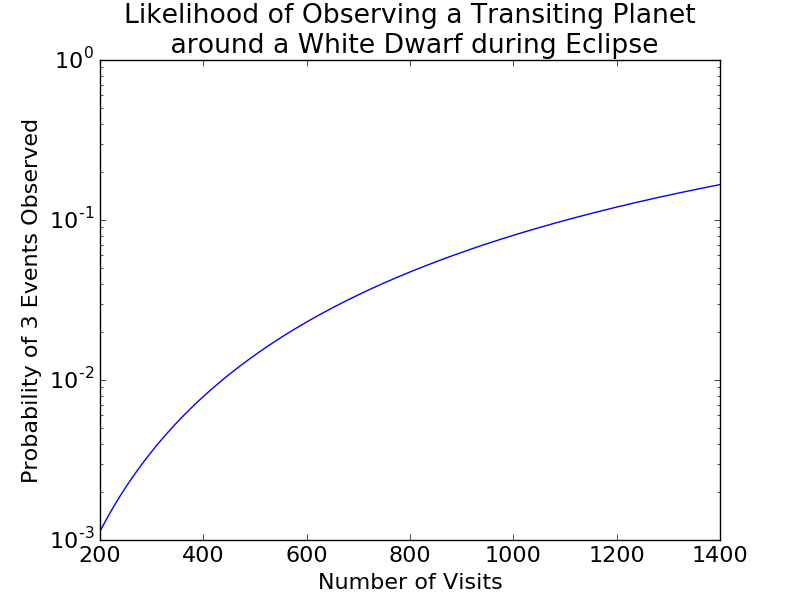}
    \caption{The chance of observing a white dwarf with a planet on a transiting orbit during the planetary eclipse at least 3 times. The Galactic plane currently recieves $\sim250$ observations, while the Wide-Fast-Deep cadence is $\sim900$ observations.}
    \label{fig:WD_observations}
\end{figure}
\clearpage

\vspace{.6in}

\section{Technical Description}
\begin{footnotesize}
{\it Describe your survey strategy modifications or proposed observations. Please comment on each observing constraint
below, including the technical motivation behind any constraints. Where relevant, indicate
if the constraint applies to all requested observations or a specific subset. Please note which 
constraints are not relevant or important for your science goals.}
\end{footnotesize}

\subsection{High-level description}
\begin{footnotesize}
{\it Describe or illustrate your ideal sequence of observations.}
\end{footnotesize}

Currently, the Galactic plane receives only $\sim250$ observations across all six bands. We would like to emphasize the importance of the Galactic plane for variable science due to the high stellar density, and observe the Galactic plane at least $\sim750$ times, with a focus primarily in only \emph{gri} bands, or some subset thereof.

\vspace{.3in}

\subsection{Footprint -- pointings, regions and/or constraints}
\begin{footnotesize}{\it Describe the specific pointings or general region (RA/Dec, Galactic longitude/latitude or 
Ecliptic longitude/latitude) for the observations. Please describe any additional requirements, especially if there
are no specific constraints on the pointings (e.g. stellar density, Galactic dust extinction).}
\end{footnotesize}

The Galactic plane subsurvey is currently defined as a region with corners at $\pm85^{\circ}$ in Galactic longitude and $\pm10^{\circ}$ in Galactic latitude. This subsurvey would attempt to observe as much of that footprint as possible. Some proposals have suggested removing areas of high extinction outside this region from the wide-fast-deep survey and grouping them with the Galactic plane, and we would want to include those regions in this proposal as well.

\subsection{Image quality}
\begin{footnotesize}{\it Constraints on the image quality (seeing).}\end{footnotesize}

We have no strong constrains on image quality. Even though we are observing in the Galactic plane, which is dense, and assuming bad seeing conditions, LSST should still do better at resolving blends than any other dedicated transit survey. Blended stars in the dense fields of the Galactic plane might influence only the measured depth. However, methods exist to entangle these blends; one example of the power of difference imaging to provide photometry in crowded fields already in use is the light curves that can be obtained from TESS using difference imaging pipelines \citep{Oelkers2018}.

\subsection{Individual image depth and/or sky brightness}
\begin{footnotesize}{\it Constraints on the sky brightness in each image and/or individual image depth for point sources.
Please differentiate between motivation for a desired sky brightness or individual image depth (as 
calculated for point sources). Please provide sky brightness or image depth constraints per filter.}
\end{footnotesize}

We are interested in Galactic objects on the bright end of LSST's capabilities, therefore we have no strong constrains on sky brightness or individual image depth. Sky brightness might only slightly increase the noise in our light curves.

\subsection{Co-added image depth and/or total number of visits}
\begin{footnotesize}{\it  Constraints on the total co-added depth and/or total number of visits.
Please differentiate between motivations for a given co-added depth and total number of visits. 
Please provide desired co-added depth and/or total number of visits per filter, if relevant.}
\end{footnotesize}

As the scientific goals outlined are for variable science, the co-added image depth is not a priority for this proposal. However, the science capable would be minimal when the number of visits is less than $\sim500$ visits.

\subsection{Number of visits within a night}
\begin{footnotesize}{\it Constraints on the number of exposures (or visits) in a night, especially if considering sequences of visits.  }
\end{footnotesize}

We do not put any constraints on the number of visits within a night.

\subsection{Distribution of visits over time}
\begin{footnotesize}{\it Constraints on the timing of visits --- within a night, between nights, between seasons or
between years (which could be relevant for rolling cadence choices in the WideFastDeep. 
Please describe optimum visit timing as well as acceptable limits on visit timing, and options in
case of missed visits (due to weather, etc.). If this timing should include particular sequences
of filters, please describe.}
\end{footnotesize}

As we hope to have well-sampled light curves in phase space for objects at a range of periods, our only constraint on how visits are distributed is that there should not be any strong aliasing present, as any such aliasing will make it more difficult to find true periods. For a more thorough explanation of how to minimize aliasing, see the complementary white paper Bell et al.

\subsection{Filter choice}
\begin{footnotesize}
{\it Please describe any filter constraints not included above.}
\end{footnotesize}

Transiting events are, to a first-order, achromatic events. As such, we only need precise photometry without being strictly band-dependent. For our objects, the best photometry will be available in \emph{gri}. It would be further beneficial to have a significant fraction of observations in a second band so that event depth can be compared in two colors to confirm that the signal appears achromatic.

\subsection{Exposure constraints}
\begin{footnotesize}
{\it Describe any constraints on the minimum or maximum exposure time per visit required (or alternatively, saturation limits).
Please comment on any constraints on the number of exposures in a visit.}
\end{footnotesize}

Concerns with saturation lead us to prefer 2 15-second exposures over 1 30-second exposure. Brighter variables sources will be better suited to follow-up observations and can add to the observations from prior surveys to increase the observed baseline.

Silvotti et al. also has highlighted that for short duration events (such as transits of white dwarfs), two 15-second exposures will help to verify that a signal is real, and minimize photometric dilution of very short-duration events.

\subsection{Other constraints}
\begin{footnotesize}
{\it Any other constraints.}
\end{footnotesize}

N/a

\subsection{Estimated time requirement}
\begin{footnotesize}
{\it Approximate total time requested for these observations, using the guidelines available at \url{https://github.com/lsst-pst/survey_strategy_wp}.}
\end{footnotesize}
Each visit would require:
\begin{itemize}
	\item Slew to field (mean 6.8 second slew time \footnote{Time estimate from \url{https://www.lsst.org/scientists/keynumbers}})
	\item First 15-second exposure (16 seconds including shutter times)
	\item 2-second readout
	\item Second 15-second exposure (16 seconds including shutter times)
\end{itemize}
For each visit, this represents 34 seconds on a given field, plus 6.8 seconds of slew time under the stated efficiencies, for a total of 41.8 seconds per pointing.

The currently defined Galactic Plane consists of 177 fields that to be treated as Wide-Fast-Deep-like fields, as per simulated cadence astro\_lsst\_01\_1004 \citep{Marshall2017}.

At 750 observations per field, this represents 8.7 hours per field, or a total of 1541 hours for this Galactic Plane region. This represents 6.4\% of the total time (3650 nights x 0.83 uptime x 8 hours/night = 24236 hours) available.
\vspace{.3in}

\begin{table}[ht]
    \centering
    \begin{tabular}{l|l|l|l}
        \toprule
        Properties & Importance \hspace{.3in} \\
        \midrule
        Image quality & 2    \\
        Sky brightness & 3 \\
        Individual image depth & 2  \\
        Co-added image depth & 3  \\
        Number of exposures in a visit   & 2  \\
        Number of visits (in a night)  & 3  \\ 
        Total number of visits & 1  \\
        Time between visits (in a night) & 3 \\
        Time between visits (between nights)  & 3  \\
        Long-term gaps between visits &  3 \\
        Other (please add other constraints as needed) & \\
        \bottomrule
    \end{tabular}
    \caption{{\bf Constraint Rankings:} Summary of the relative importance of various survey strategy constraints. Please rank the importance of each of these considerations, from 1=very important, 2=somewhat important, 3=not important. If a given constraint depends on other parameters in the table, but these other parameters are not important in themselves, please only mark the final constraint as important. For example, individual image depth depends on image quality, sky brightness, and number of exposures in a visit; if your science depends on the individual image depth but not directly on the other parameters, individual image depth would be `1' and the other parameters could be marked as `3', giving us the most flexibility when determining the composition of a visit, for example.}
        \label{tab:obs_constraints}
\end{table}

\subsection{Technical trades}

%\begin{footnotesize}
%{\it To aid in attempts to combine this proposed survey modification with others, please address the following questions:}
%\end{footnotesize}
\begin{enumerate}
    \item What is the effect of a trade-off between your requested survey footprint (area) and requested co-added depth or number of visits?
    
    Footprint (and more specifically, number of stars within the footprint) has a significant trade-off with the number of visits. For a fixed amount of survey time, at least in the white dwarf case, increasing the number of observations improves the science possible much more than the loss in number of stars observed will harm it.
    \item If not requesting a specific timing of visits, what is the effect of a trade-off between the uniformity of observations and the frequency of observations in time? e.g. a `rolling cadence' increases the frequency of visits during a short time period at the cost of fewer visits the rest of the time, making the overall sampling less uniform.
    
    These science goals can be achieved equally well with or without a rolling cadence, provided that aliasing is minimized in the observations.
    \item What is the effect of a trade-off on the exposure time and number of visits (e.g. increasing the individual image depth but decreasing the overall number of visits)?
    
    As the priority of this work is variability, increasing individual depth while decreasing the number of visits will be a major limitation to the possible science as well-sampled phase coverage is the most important requirement.
    \item What is the effect of a trade-off between uniformity in number of visits and co-added depth? Is there any benefit to real-time exposure time optimization to obtain nearly constant single-visit limiting depth?
    
    There is no major benefit to real-time exposure time optimization.
    \item Are there any other potential trade-offs to consider when attempting to balance this proposal with others which may have similar but slightly different requests?
    
    There are at least two similar proposals that have been made that are more broadly aimed at increasing the number of observations in the Galactic plane (Street et al and Strader et al). However, while these proposals may argue for an increase of observations in the whole plane with a prioritization on the footprint, our proposal is to instead focus on observing as much of the plane as possible at a suitably high cadence for transiting planet detection, with the priority on first reaching an observing threshold.
\end{enumerate}

\section{Performance Evaluation}
The performance of a survey in the plane that is capable of detecting periodically-occurring events will depend most strongly on how many observations are made during these recurring events, but without the constraint that these points must take place during a single event. This means that the largest factor for detection is simply the total number of observations.

There are two metrics that can be used to assess this. The first is a figure of merit that is created by combining two smaller metrics. The first is the stellar occurrence rate. This can be sourced from the stellarDensityMap that is included in MAF (which can select for only white dwarfs), or the StarCountMetric that is part of sims\_maf\_contrib. The second component is a newly-written metric, EventFractionMetric. The design of this metric is to provide a first-order approximation for how likely it is to observe a periodically-repeating event above some threshold with a set of observations that randomly sample phase space. For example, the likelihood of observing a transiting Earth-sized planet around a white dwarf during the transit itself is on order of 0.001 and \citet{Lund2018} suggests that at least 3 observations are necessary. At 250 observations, the chance of this occurring is 0.002. At 500 observations, this increases to 0.014, and at 750 observations reaches 0.04, showing that increasing the number of observations by a factor of 3 increases potential candidates by a factor of 20. For transits around main-sequence stars, the fractional transit duration is longer ($\sim0.05$ for a Hot Jupiter around a solar-mass star) but with shallower events additional points in transit will be needed for detection, likely on the order of tens of points in transit. The product of the EventFractionMetric and a metric for stellar population gives, to first order, a measurement of how many stars can be searched for planetary transits.

The second metric is complimentary to the above, and is a diagnostic metric for reducing aliasing. Also in sims\_maf\_contrib is the Periodogram Purity Function (PeriodicMetric), which uses the window function of scheduled observations to determine how much strength of a periodic signal is being lost due to aliasing \citep{Lund2016}. In this metric, each point is given a value between 0 and 1, with 1 being an ideal sampling with no impact from aliasing. A more thorough proposal to address aliasing can be found in the aliasing white paper by Bell et al.

There are significant similarities in this proposal and Street et al.'s proposal for increased observations of the Galactic Plane, however they differ in the metrics that are most concerned with. In this proposal, we are most concerned with the overall number of observations while minimizing aliasing. In contrast, the microlensing science in Street et al. places a higher priority on uniform spacing in the cadence. We believe, however, that these two proposals can be merged to result in a survey that will be well-suited for the science cases in both proposals.

%\begin{footnotesize}
%{\it Please describe how to evaluate the performance of a given survey in achieving your desired science goals, ideally as a heuristic tied directly to the observing strategy (e.g. number of visits obtained within a window of time with a specified set of filters) with a clear link to the resulting effect on science. More complex metrics which more directly evaluate science output (e.g. number of eclipsing binaries successfully identified as a result of a given survey) are also encouraged, preferably as a secondary metric.If possible, provide threshold values for these metrics at which point your proposed science would be unsuccessful and where it reaches an ideal goal, or explain why this is not possible to quantify. While not necessary, if you have already transformed this into a MAF metric, please add a link to the code (or a PR to \href{https://github.com/lsst-nonproject/sims_maf_contrib}{sims\_maf\_contrib}) in addition to the text description. (Limit: 2 pages).}
%\end{footnotesize}

\vspace{.6in}

\section{Special Data Processing}
\begin{footnotesize}
{\it Describe any data processing requirements beyond the standard LSST Data Management pipelines and how these will be achieved.}
\end{footnotesize}

We do not anticipate any special processing requirements.

\section{Acknowledgements}
This work developed partly within the TVS Science Collaboration and the authors acknowledge the support of TVS in the preparation of this paper.

The authors acknowledge support by the FlatIron Institute and Heising-Simons Foundation for the development of this paper.

\section{References}
\bibliographystyle{apalike}
\bibliography{refs}
\end{document}